# Magnetoelectric Effect driven by Magnetic Domain Modification in LuFe$_2$O$_4$

Takashi Kambe[1*], Yukimasa Fukada[1], Jun Kano[1], Tomoko Nagata[1], Hiroyuki Okazaki[1], Takayoshi Yokoya[2], Shuichi Wakimoto[3], Kazuhisa Kakurai[3] and Naoshi Ikeda[1]

[1]*Department of Physics, Faculty of Science, Okayama University, Okayama 700-8530, Japan*
[2]*Research Laboratory for Surface Science, Okayama University, Okayama 700-8530, Japan*
[3]*Quantum Beam Science Directorate, Japan Atomic Energy Agency, 2-4 Shirakata Shirane, Tokai, Naka, Ibaraki 319-1195, Japan,*

Magneto-capacitance effect was investigated using the impedance spectroscopy on single crystals of LuFe$_2$O$_4$. The intrinsic impedance response could be separated from the interfacial response and showed a clear hysteresis loop below $T_{\text{Ferri}} \sim 240$ K under the magnetic field. The neutron diffraction experiment under the magnetic field proves the origin of dielectric property related to the motion of nano-sized ferromagnetic domain boundary. These results imply that the modification of the microscopic domain structure is responsible for the magnetoelectric effect in LuFe$_2$O$_4$.

* Corresponding author; kambe@science.okayama-u.ac.jp

The strong coupling between charge and spin ordering dominates magneto-electric (ME) phenomena [1]. The ME effect has lately attracted considerable attention and is particularly noticed in *multiferroic* materials [1,2]. $R$Fe$_2$O$_4$ (RFO, $R$; rare earth ions) is one of the candidate materials for the ME phenomenon. The colossal dielectric properties have been reported in RFO [3,4,5]. It is insisted that the dielectric response in RFO originates in the dynamical electronic exchange between $Fe^{2+}$ and $Fe^{3+}$ [6]. That is, the motion of a domain boundary proceeds by electron exchange between the iron ions. However, recently, extrinsic contributions from grain boundaries and electrode contacts have been pointed out in polycrystalline [7] and single crystal LuFe$_2$O$_4$ (LFO) [8]. The extrinsic dielectric properties have also been debated for many oxides [9-12], where a leaky component and electric polarization coexist. Accordingly, a further study using single crystals is quite important to extract the intrinsic dielectric permittivity and ME effect of RFO.

The crystal structure of RFO consists of layered triangular lattices of Fe ions [3]. Doubly stacked triangular layers of Fe ions make a unit (abbreviated as *W*-layer). As equal amounts of $Fe^{2+}$ and $Fe^{3+}$ ions coexist in the *W*-layer, the average valence of Fe ion is +2.5. Geometrical frustration on the triangular lattice plays a key role in selecting the ground state configuration for the charge as well as the spin [13]. A three-dimensional ordering of the Fe valences takes place below $T_{CO}$ ~ 330 K. Since the Fe charge-ordering (CO) pattern in the *W*-layer has no inversion symmetry, it has been suggested that a spontaneous ferroelectric polarization originate in coherent CO, a group now categorized as "*electronic ferroelectric materials*" [3,14]. Moreover, the antiferromagnetic interactions among $Fe^{2+}$ and $Fe^{3+}$ ions leads to a ferrimagnetic ordering below $T_{Ferri}$ ~ 240 K [15]. In this Letter, we report the investigation of ME response in a single crystal LFO with different electrodes. We found strong contact effects between the metallic electrode and the semiconducting LFO and succeeded in extracting the intrinsic permittivity and resistivity by selecting appropriate electrodes. We will present the ME response under magnetic field, and will also show the neutron diffraction experiments under the magnetic field to clarify the microscopic magnetic domain structure.

First, we show the contact effects on the impedance response. The upper parts of Fig. 1 show the frequency dependence of capacitance and resistance for single crystals of LFO and the lower parts of Fig. 1 show Z'- Z'' plots at 220 K (below $T_{Ferri}$), where Z' and Z'' denote the real and imaginary parts of impedance, respectively. The electric field was applied parallel to the *c*-axis. With the Ag electrode, the capacitance exhibits a step-like increase at low frequencies around $10^3$ Hz in addition to a small step at high frequencies around $10^4$ Hz (see the inset of Fig.1). The Z'-Z'' plot for the Ag electrode exhibits more pronounced behavior and is clearly separated into two semicircular components. These characteristic impedance properties can be well reproduced by an equivalent circuit with two CR parallel circuits as shown in the inset of Fig. 1. The semicircle at the high Z' side (low frequency side) corresponds to a contribution from circuit 1, while that at the low Z' side (high frequency side) corresponds to that from circuit 2. Hereinafter, we define these responses as the high-Z' and the low-Z', respectively. The impedance responses in all investigated samples are decomposed into two components. Interestingly, while the high-Z' response strongly depends on the electrode material, the low-Z' response depends only slightly on them. If we apply the equivalent circuit for three electrodes, almost identical parameters of $R_s$ and $C_s$ can be investigated from the low-Z' components. The small difference in the low-Z' response may be associated with differences in the sample size and/or the temperature for each measurement. Accordingly, the low-Z' response should reflect the bulk properties of LFO though an extrinsic interfacial effect between the sample and the electrode may give rise to the high-Z' response.

Figure 2 (a) shows the temperature dependence of $R_s$ and $C_s$ for the Au electrode. The high-Z' response exhibits a negligibly small contribution. We obtain an extremely small $C_s$ component for the low-Z' response. Using the simple relation, $C = \varepsilon_0 \varepsilon S / d$, the relative permittivity, $\varepsilon$, can be estimated to be about 40. The value of $\varepsilon$ increased weakly with increasing temperature and a small peak seems to appear around $T_{\text{Ferri}}$. Unfortunately, above 280 K, it is difficult to obtain a response that permits the determination of $\varepsilon$ due to the increase in conductivity. With the Ag electrode, however, if we estimate $\varepsilon$ from the total capacitance without decomposing the two contributions, we obtain a two-orders-of-magnitude larger $\varepsilon$, which is comparable with previous experiments [3,4,5].

Figure 2 (b) shows the Arrhenius plot of the relaxation frequency, $f_2$, for the low-Z' response. The value of $f_2$ is clearly described by the activation relation, $f_2 \sim 1/R_s C_s \exp(-E_g/k_B T)$, with an energy gap of $E_g = 335$ meV for $T > 190$ K, and $= 146$ meV for $T < 190$ K. The average $E_g = 240$ meV is consistent with the transport and Mössbauer experiments [6]. By extrapolating this relation to higher temperature, we estimate $f_2 \sim 3 \times 10^6$ Hz at room temperature. This value also corresponds well to the relaxation frequency estimated by the Mössbauer experiments [6,16], which have proved the observation of the charge fluctuation between $Fe^{2+}$ and $Fe^{3+}$ ion. The temperature of 190 K is near the structural phase transition temperature $T_{\text{LT}} \sim 170$ K [16,17]. In Fig. 2 (c), we plot both the resistivity measured by the impedance method, $\rho_{\text{im}}$, and by the usual DC method, $\rho_{\text{dc}}$, with the Au electrode. The temperature dependences of $\rho_{im}$ and $\rho_{dc}$ are completely consistent with each other. Thus, the $\rho_{im}$ or the low-Z' response is essentially associated with the bulk properties of LFO. The temperature dependence of the resistivity can be divided in two regions below $T_{CO}$. From $T_{CO}$ to 200 K, both $\rho_{im}$ and $\rho_{dc}$ yield activation relations with $E_g$ = 360 meV, while below 200 K, $\rho_{im}$ has $E_g$ = 155 meV. These values are completely consistent with the $E_g$ derived from the temperature dependence of the $f_2$.

Furthermore, we note that the $\rho_{dc}$ is also affected by the species of electrode. Figure 2 (d) shows the temperature dependence of DC resistances with the Ag and Au electrodes. With the Ag electrode, the resistance increased by fully two orders of magnitude even at room temperature, and the anomaly around $T_{CO}$ completely disappeared. The black, blue and red crosses represent the R$s$, the R$i$ and the series (R$s$ + R$i$) resistances, respectively, estimated by the impedance method with the Ag electrode at 220 K. It is reasonable that the R$s$ component with the Ag electrode almost corresponds to the resistance with the Au electrode, implying the intrinsic bulk resistance. Although we could not measure the resistance with the Ag electrode below 230 K, we notice that the series resistance at 220 K follows as an extension of the R-$T^{-1}$ curve measured by the DC resistance method. On the contrary, with the Au electrode, we conclude that the intrinsic resistivity of LFO can be directly investigated without taking interfacial effects into consideration.

The high-Z' response strongly depends on the species of metallic electrode, with their contributions possibly weakening in the order of the magnitude of the work function, $\phi_M$, of the metallic electrode. The $\phi_M$s for Ag, C, and Au are known to be 4.6, 5.0 and 5.2 eV, respectively. Here, we propose that the formation of a Schottky barrier junction close to the metallic electrode leads to the high-Z' response. The band structure of YbFe$_2$O$_4$ (YbFO) was investigated by photo-emission (PES) and inverse photo-emission spectroscopy (IPES) [18]. We estimated an ionic potential of ~ 6 eV, an electron affinity of ~ 4 eV and a Fermi energy, $\varepsilon_F$, of ~ 5.2 eV. In Fig. 3, the band structure of YbFO and the $\phi_M$s of electrodes are shown together with the spectra from PES/IPES for YbFO. Since the band structure of the RFO family near the $\varepsilon_F$ should be identical, the $\varepsilon_F$ of LFO should be located around ~ 5.2 eV, which is very close to the $\varepsilon_F$ of Au. If no Schottky barrier forms at the interface between Au and LFO, we expect no interfacial effects would be activated. Accordingly, these results suggest that the formation of the high Schottky barrier at the interface may yield a colossal dielectric permittivity of LFO.

Next, we focus on the magnetic and electric field dependence in the impedance response of LFO. Figures 4 (a) and (b) show the Z'-Z'' plot for the impedance response with the carbon electrode as a function of electric and magnetic field, respectively. To investigate which contribution was responsible for the ME coupling, we analyzed the Z'-Z'' response with the carbon electrode under a magnetic/electric field because two moderate contributions are observed.

As shown in Fig. 4 (a), when the AC electric field is increased, the radius of the high-Z' response is clearly diminished while no effect on the low-Z' response can be detected. Thus, the electric field can selectively modify the interfacial impedance. Conversely, we notice that the impedance of sample is kept within the limit of linear response. Thus, the non-linear current-voltage response [19] may be reconsidered as the interfacial effect. The width of the depletion layer increases with increasing applied electric field, leading to a decrease in interfacial capacitance. The $C_i$ component obtained from the high-Z' response decreases slightly with increasing electric field, which may qualitatively explain the change in the high-Z' response. However, the shrinkage of radius in the Z'-Z'' plot directly implies a decrease in the resistance component of the circuit. Further study, which includes the DC-bias effects on the interface between the metal and LFO, will be needed to clarify the electric field effect on the interfacial impedance.

In Fig. 4 (b), the Z'-Z'' plots with a carbon electrode are shown as a function of the magnetic field at 220 K, where both the magnetic field and the electric field are applied parallel to the *c*-axis. When the magnetic field is increased, the low-Z' response is mainly modified while the high-Z' response shows a negligibly small change. So, the magnetic field has an effect on the low-Z' response, *i.e.*, the bulk properties of the sample, which is contrary to the case for the electric field. In Fig. 4 (c), we summarize the magnetic field dependence of estimated parameters with the Au electrode and the magnetic field dependence of magnetization along the *c*-axis at 220 K (note that the high-Z' response can be neglected for the Au electrode). Below $T_{Ferri}$, the magnetization traces out a clear hysteresis loop and all the parameters also trace out clear hysteresis loops. It should be noted that the electron hopping frequency between $Fe^{2+}$ and $Fe^{3+}$ is enhanced by the application of the magnetic field. The magneto-resistance plays a dominant role in the magnetic field dependence of the low-Z' response, and the resistance is strongly suppressed by the magnetic field. The observed magneto-capacitance (MC) effect, *i.e.* the ME effect, is quite weak as compared with the previous report [4], in which we may consider the interfacial MC effects have been included. Note that this weak MC effect was not definitely observed until the interfacial component was removed, *i.e.*, until, concretely, we used the Au electrode.

The alignment of the ferrimagnetic moments [20] along the *c*-axis leads to possible domain structures. It was proposed in LFO that the macroscopic magnetization originated from a sum of the distributed ferrimagnetic microscopic domains [21]. The magnetic hysteresis loop observed in low-magnetic field may give rise to the change in the volume fraction of competing domains. In order to directly probe the microscopic magnetic structure of LFO, we performed the neutron diffraction experiments. The reciprocal space traces were done on (1/3, 1/3, -2) and (1/3, 1/3, -2,5) peaks to estimate the coherence length ($\xi$) of the magnetic domain as a function of magnetic field. These magnetic peaks inform the ferromagnetically and antiferromagnetically ordered ferrimagnetic domains, respectively.

As shown in Fig. 5, the half-width of half-maximum (HWHM) of (1/3, 1/3, -2) peak along the *l*-direction decreases with increasing magnetic field, and shows a clear hysteresis loop, where the magnetic field is applied parallel to the *c*-axis. Considering the experimental resolution, we estimate the $\xi$ along the *c*-axis, $\xi_c(0)$, ~ 100 Å at zero magnetic field. On the contrary, the (1/3, 1/3, -2) peak along the *l*-direction at 0.28 Tesla has a resolution-limited profile, suggestive of sufficiently long $\xi_c$ (> 500 Å). The size of the $\xi_c(0)$ is consistent with the size of the spin and the CO domain along the *c*-axis obtained previously [22]. The peak intensity also increases under the magnetic field and saturates around 0.28 Tesla. These results indicate that the relative volume of the ferromagnetic domain develops with increasing the magnetic field while that of other domain suppresses.

The $\xi$ within the *ab*-plane, $\xi_{ab}$, can be estimated from the scan along the *h*-direction and is obtained to be ~ 40 Å at zero magnetic field. This value of the $\xi_{ab}$ is one-order smaller than the size of CO domain within the *ab*-plane obtained previously [22]. At 0.28 Tesla, the $\xi_{ab}$ slightly increases (~ 60 Å) and the intensity also weakly increase. Accordingly, a nano-sized rod-like ferromagnetic domain (~100 Å along the *c*-axis, ~ 40 Å in the *ab*-plane) is formed around zero magnetic field. Under the magnetic field along the *c*-axis, the ferromagnetic domain is mainly extended along the *c*-axis and, finally, the long rod-like ferromagnetic domain (>500 Å along the *c*-axis, ~ 60 Å in the *ab*-plane) is formed at 0.28 Tesla. The neutron diffraction experiments clearly demonstrate that the hysteresis on the bulk physical properties originates from the change in the magnetic domain size or pattern and imply the existence of strong pinning effects for the magnetic domain. That is, the balance of volume fraction for the magnetic domains at zero magnetic field should be maintained even if the samples are experienced in the high-magnetic field. It is claimed that the CO domain boundaries act as the pinning center for the magnetic domain [22]. The application of the magnetic field along the *c*-axis produces a flop for the ferrimagnetic component at the magnetic domain boundary. The observed MC effect suggests that this spin-flop leads to the modulation of the electronic charge on Fe ions at the domain boundary, which may be explained by the charge-modulated spin-exchange mechanism [23].

In summary, the ME effect on single crystals of LFO was evidenced by removing the extrinsic interfacial component from the whole impedance response. The neutron diffraction experiment proved the origin of dielectric property related to the motion of nano-sized ferromagnetic domain boundary. The present study should provide the basis for better insight into the electronic ferroelectricity of RFO.

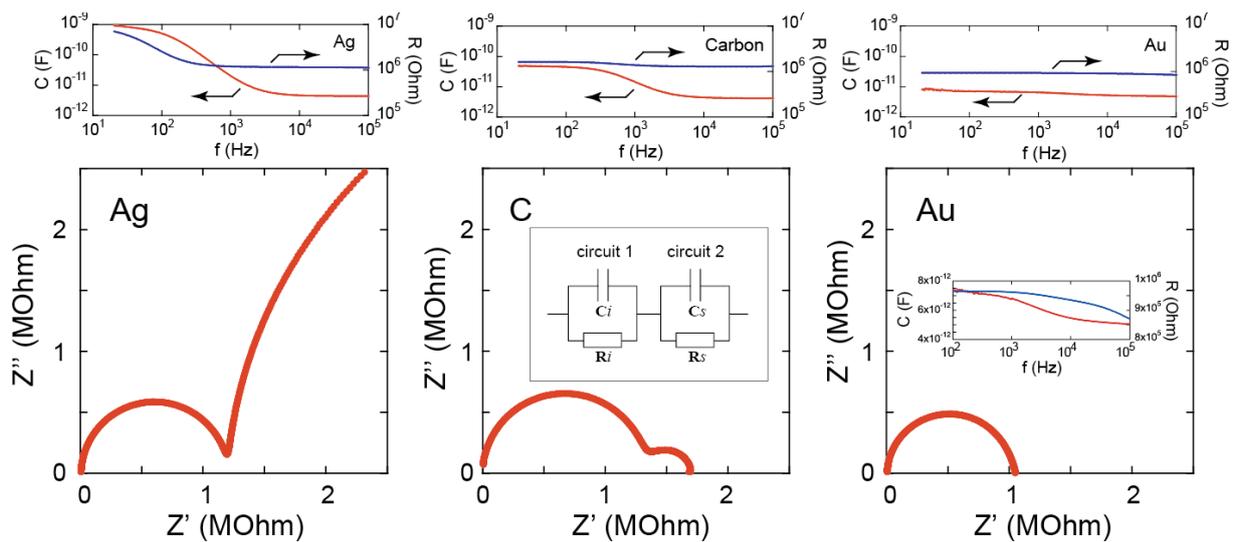

Figure 1.

Impedance spectra (upper part) and Z'-Z'' plots (lower part) at 220 K using three different electrodes (Ag, Carbon and Au). Z' and Z'' denote the real and imaginary parts of impedance, respectively. The inset figure in the Z'-Z'' plot with the carbon electrode shows an equivalent circuit with two CR parallel circuits. The inset figure in the Z'-Z'' plot with the Au electrode shows the enlargement of the impedance spectra.

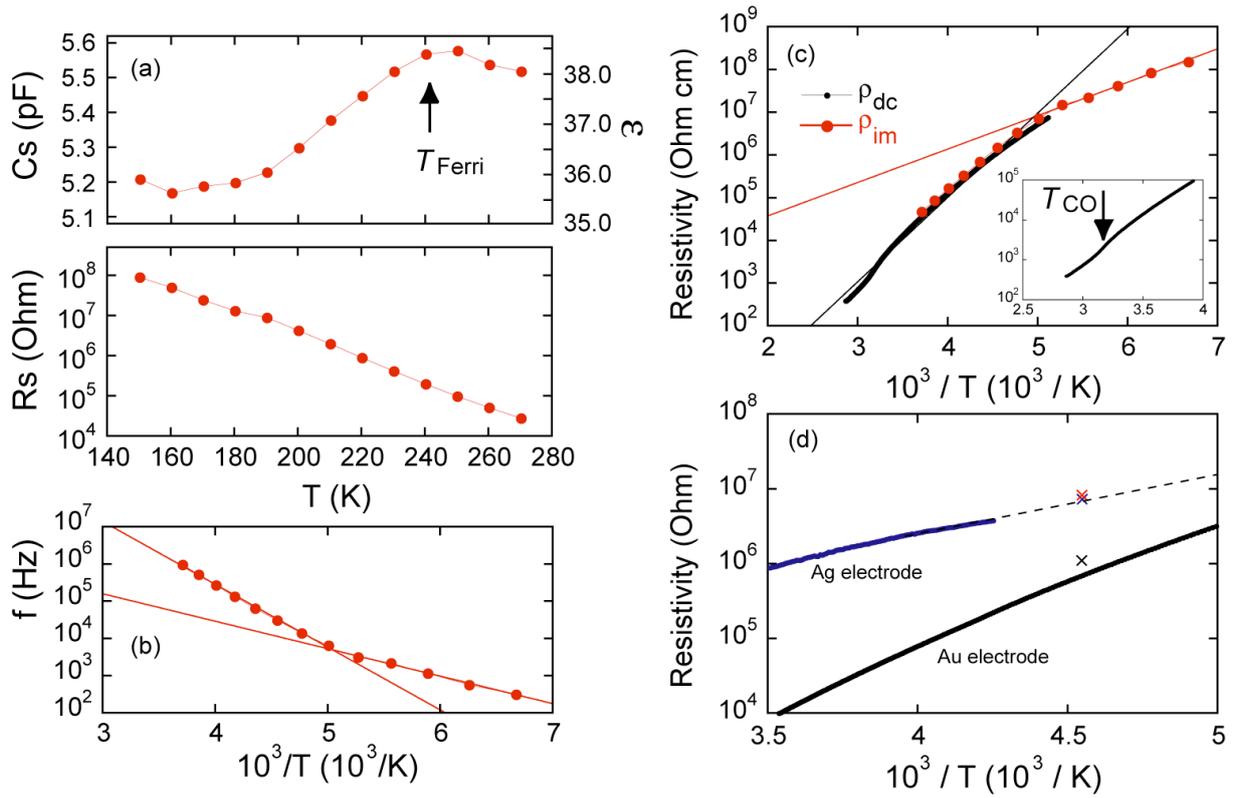

Figure 2.

(a) Temperature dependence of capacitance ($C_s$) and resistance ($R_s$) with the Au electrode. The right-hand scale in the $C_s$ plot denotes the dielectric permittivity, $\varepsilon$. (b) Arrhenius plot of relaxation frequency with the Au electrode. (c) Arrhenius plot of resistivity with the Au electrode. The resistivity measured by the impedance method, $\rho_{im}$, and by the DC method, $\rho_{dc}$, are plotted. The inset shows the enlargement around $T_{CO}$. (d) Temperature dependences of the DC-resistances with the Ag and Au electrodes. The black, blue and red crosses represent the sample ($R_s$), the interfacial ($R_i$) and total ($R_s + R_i$) resistance, respectively, estimated by the impedance method with the Ag electrode at 220 K. Dashed line is a guide for the eye.

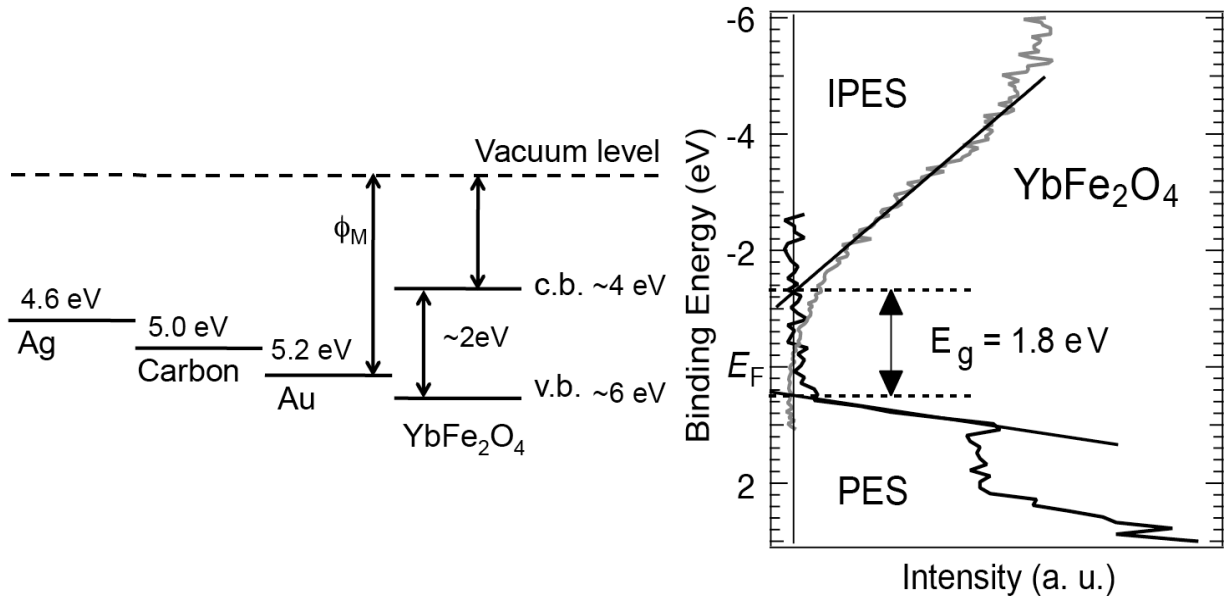

Figure 3.

Band structure of YbFO and the work function, $\phi_M$, of the electrodes are shown together with the PES and the IPES spectra for YbFO [14].

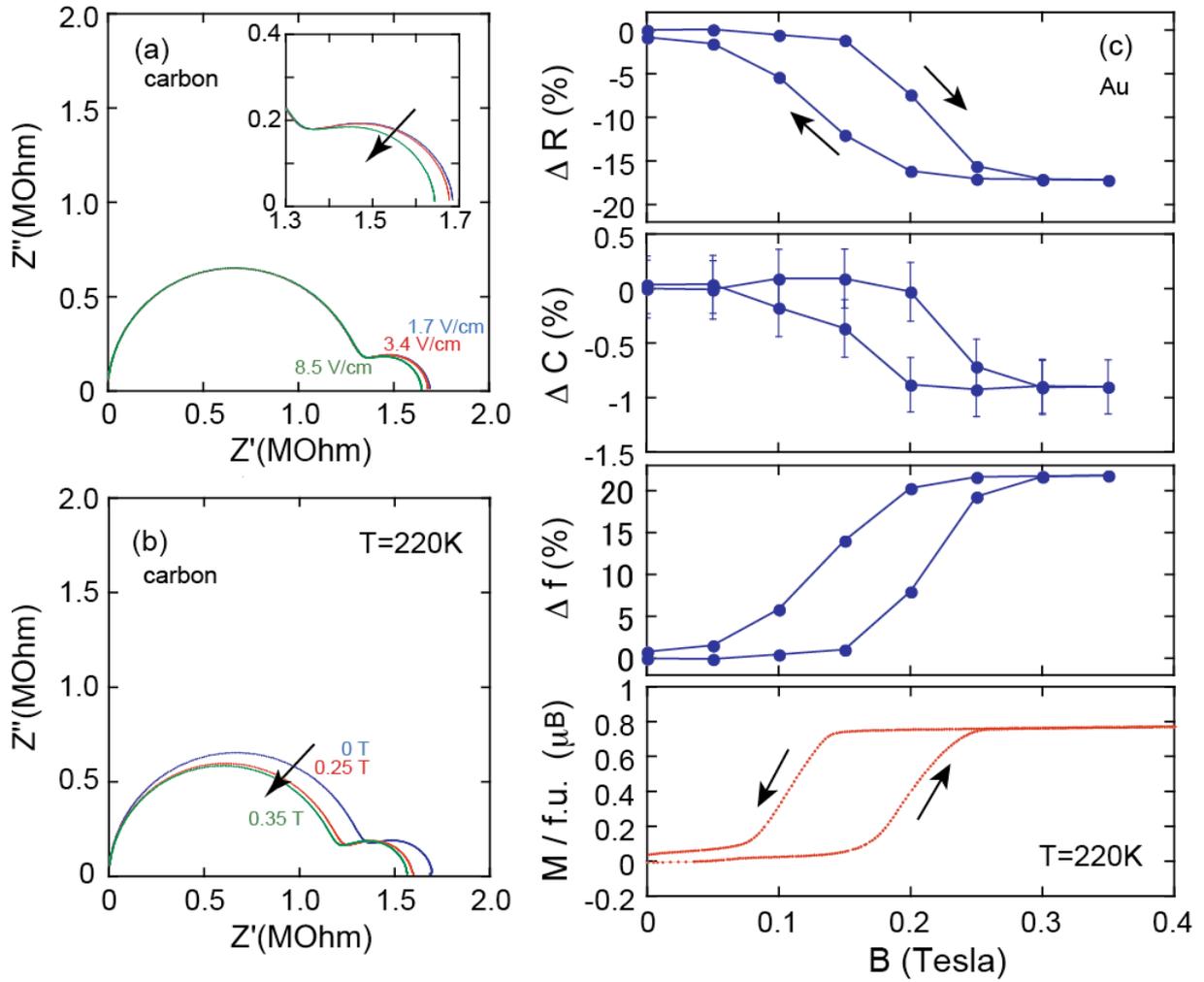

Figure 4.

Z'-Z'' plots with the carbon electrode as a function of electric field (a) and magnetic field (b) at 220 K. (c) Magnetic field dependence of resistivity, capacitance and relaxation frequency estimated from impedance spectroscopy with the Au electrode: $\Delta R = (Rs(H) - Rs(0))/Rs(0)$, $\Delta C = (Cs(H) - Cs(0))/Cs(0)$ and $\Delta f = (fs(H) - fs(0))/fs(0)$. The bottom figure shows the magnetization hysteresis curve along the *c*-axis.

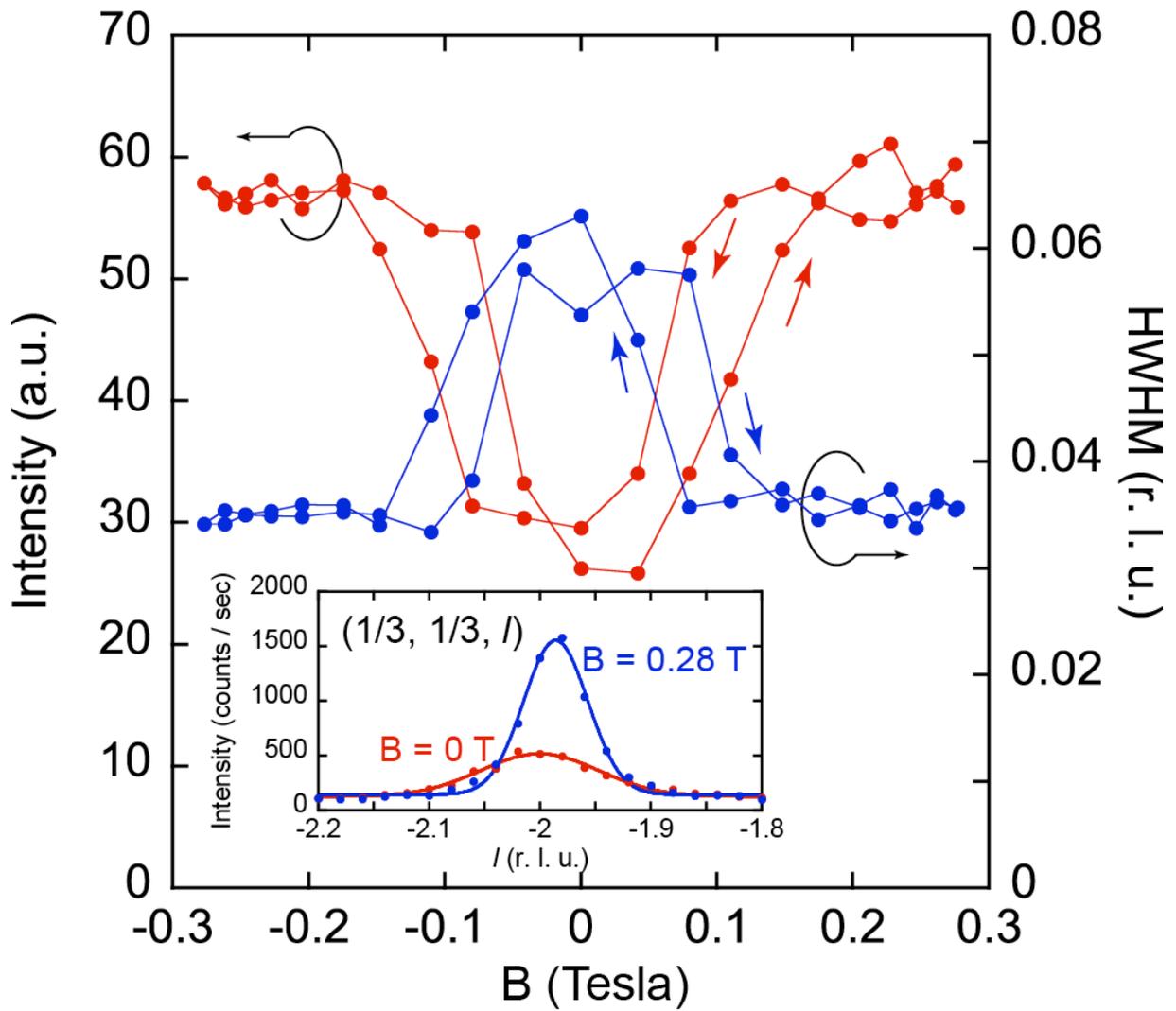

Figure 5.

Magnetic field dependence of the intensity and the half-width of half-maximum (HWHM) of the $\left(\frac{1}{3},\frac{1}{3},-2\right)$ magnetic diffraction peak. The magnetic field is applied parallel to the *c*-axis. Inset: neutron diffraction profiles along the *l*-direction at 0 Tesla (red) and 0.28 Tesla (blue). The solid lines denote the fitted results with Gaussian line shape.